\documentclass{optica-article}

\journal{opticajournal} % for journals or Optica Open

\articletype{Research Article}

\usepackage{caption}
\begin{document}

\title{Deep Learning-based Single-Shot Composite Fringe Projection Profilometry with Pixel-Wise Uncertainty Quantification}

\author{Xiangjun Kong,\authormark{1,*} Qingkang Bao,\authormark{2} Tibebe Yalew,\authormark{1} Gerardo Adesso\authormark{3} and Samanta Piano,\authormark{1,*}}

\address{\authormark{1}Manufacturing Metrology Team, Faculty of Engineering, University of Nottingham, Nottingham, UK\\
\authormark{2}School of Mechanical Engineering, Xi’an Jiaotong University, Xi’an, Shaanxi, China\\
\authormark{3}Centre for the Mathematical and Theoretical Physics of Quantum Non-Equilibrium Systems and School of Mathematical Sciences, University of Nottingham, Nottingham, UK}

\email{\authormark{*}xiangjun.kong@nottingham.ac.uk \authormark{*}samanta.piano@nottingham.ac.uk} %% email address is required; see note below about the corresponding author designation

% use {asbstract*} to suppress the copyright line. Copyright information will be added in production

\begin{abstract*}
Driven by the growing demand for high-speed 3D measurement in advanced manufacturing, optical metrology algorithms must deliver high accuracy and robustness under dynamic conditions. Fringe projection profilometry (FPP) offers high precision, yet the 2\(\pi\)  ambiguity of wrapped phase means that conventional absolute phase recovery typically relies on multiple coded patterns, sacrificing temporal resolution. Deep learning-based composite-FPP (CFPP) shows promise for single-shot phase recovery from a composite fringe, but limited interpretability makes it difficult to assess reconstruction reliability or trace error sources in the absence of ground truth. To address this, we propose HSURE-CFPP (Heteroscedastic Snapshot-ensemble Uncertainty-aware Ratio Estimation for CFPP). HSURE-CFPP predicts the numerator-denominator “ratio” used for wrapped-phase computation with a heteroscedastic snapshot-ensemble network, enabling ultra-fast 3D imaging from a single composite fringe and producing pixel-wise uncertainty maps for confidence assessment and unreliable-region identification. Specifically, a heteroscedastic likelihood jointly estimates pixel-wise noise variance to capture data uncertainty, while a snapshot ensemble quantifies model uncertainty via the dispersion across snapshots, yielding total predictive uncertainty as an interpretable reliability measure. Experiments on static and dynamic scenes demonstrate that HSURE-CFPP achieves high-accuracy reconstruction at high speed, and that the predicted uncertainty correlates well with reconstruction errors, providing a deployable quality-assessment mechanism for deep-learning-based FPP.
\end{abstract*}

%%%%%%%%%%%%%%%%%%%%%%%%%%  body  %%%%%%%%%%%%%%%%%%%%%%%%%%
\section{Introduction}
Driven by high-speed applications such as in-process defect inspection in manufacturing, there is a growing demand for real-time 3D measurement with higher throughput without sacrificing accuracy \cite{wang2025single,zhang2016situ}. Meanwhile, assessing reconstruction reliability and tracing error sources are increasingly critical in practical deployments, particularly when ground-truth 3D data are unavailable. Against this backdrop, fringe projection profilometry (FPP) has become a widely used technique for 3D shape measurement due to its non-contact operation, high precision, and low cost \cite{geng2011structured,zuo2018phase,salvi2010state}. In FPP, 3D shape is recovered through system calibration and fringe analysis, where phase extraction and phase unwrapping are key steps \cite{zuo2018phase,salvi2010state}. Phase unwrapping is a major bottleneck for speed and accuracy because the arctangent operation yields a wrapped phase with inherent 2\(\pi\) discontinuities. Common strategies include spatial phase unwrapping (SPU) and temporal phase unwrapping (TPU) \cite{zhang2018absolute,zuo2016temporal}: TPU is typically more accurate and robust to noise and discontinuities, but requires additional projections and therefore reduces sampling efficiency \cite{zuo2018phase}. Moreover, conventional FPP largely reports deterministic height without an explicit uncertainty model, limiting error-source budgeting and traceable, risk-aware conformity decisions. In metrology, a result is incomplete without its uncertainty; for FPP, this uncertainty is strongly spatially varying due to surface-dependent imaging effects and occasional unwrapping failures, motivating pixel-wise uncertainty maps for reliability-aware 3D reconstruction and inspection.

Two techniques in FPP potentially can be adopted for fast 3D measurement, namely Fourier transform profilometry (FTP) \cite{takeda1983fourier} and phase-shifting profilometry (PSP) \cite{zhang2010superfast}. As only a single fringe pattern is used in FTP, SPU is typically adopted, but it fails for objects with spatially isolated surfaces where the fringe order cannot propagate and the absolute phase becomes ambiguous. To improve phase unwrapping accuracy, spatial frequency multiplexing (FM) using composite fringe patterns has been investigated \cite{takeda1997frequency,yue2007fourier}, where multiple sinusoids with different spatial carrier frequencies are encoded into a single pattern. However, in FTP, steep surface gradients and high-frequency content can cause spectral leakage, leading to overlap between the fundamental component and adjacent spectral components, particularly in composite fringes, which degrades the effectiveness of band-pass filtering. An improved composite fringe multiplexes two sinusoidal patterns with orthogonal carrier frequencies \cite{wu2012improved}, reducing spectral overlap and enabling more reliable TPU. 
With a high-speed camera and projector, FTP can achieve ultra-fast measurement; however, the system is costly and typically involves a trade-off between frame rate and signal-to-noise ratio. Moreover, in FTP, frequency-domain filtering cannot cleanly isolate the fundamental component due to inevitable spectral aliasing, which degrades measurement accuracy even with advanced filters (e.g., orthogonal elliptic filters) \cite{chen2014fourier}. 

PSP generally achieves higher accuracy than FTP because it uses multiple phase-shifted fringes, thereby suppressing random noise and mitigating reflectivity effects \cite{salvi2010state}. For robust decoding, PSP often adopts TPU, which comes at the expense of real-time performance. With improved hardware, kilohertz-rate digital light processing (DLP) projection mitigates object-motion-induced image artefacts in PSP \cite{gong2010ultrafast}. This suits the one-bit binary defocusing approach that forms sinusoidal fringes by projector defocus, although the resulting low image contrast can reduce phase accuracy. To avoid multi-frame projection, a single-shot composite-pattern PSP has been explored. For instance, a three-chip RGB camera captures a colour-encoded pattern in which three fringe components are multiplexed into the colour channels \cite{zhang2006time}. However, colour crosstalk and non-linear sensor response reduce the accuracy of phase extraction. In addition, a speckle-embedded fringe approach embeds a speckle-like signal within three sinusoidal phase-shifted fringes to enable absolute phase recovery \cite{zhang2013unambiguous}. Despite its speed, the method is noise sensitive and exhibits an inherent trade-off between spatial resolution and measurement accuracy. 

Recent deep-learning methods based on data-driven, especially CNNs, have been introduced into FPP to alleviate phase ambiguity and noise \cite{zuo2022deep,feng2019fringe,yin2023physics,yu2020deep,wang2025single}. They can recover absolute phase from a single frequency fringe pattern more accurately than conventional approaches (e.g., FTP) \cite{yin2023physics,yu2020deep,wang2025single}. However, directly regressing wrapped phase, fringe order, or unwrapped phase from a single frequency fringe provides limited physical constraints, increasing decoding difficulty and potentially reducing stability and accuracy. Moreover, applying SPU to the predicted wrapped phase can further degrade accuracy. Therefore, deep learning-based composite FPP (CFPP) has been explored to achieve real-time measurement while maintaining phase demodulation accuracy, since a composite fringe provides richer cues than a single frequency fringe \cite{li2025deep,chen2024deep,jiang2024deep,li2022composite,fu2024deep,zhang2025deep}. Among these, CNN-assisted colour CFPP can mitigate inter-channel crosstalk \cite{fu2024deep,zhang2025deep}, but it is unreliable on coloured surfaces where spectral reflectance distorts channel-wise fringe contrast. To avoid colour dependence, FM CFPP has been used for absolute phase retrieval \cite{li2022composite,li2025deep,chen2024deep,jiang2024deep}. However, multiplexing multiple high-frequency components into a single pattern reduces fringe contrast due to modulation transfer function (MTF) attenuation and increases spectral crosstalk when the components are closely spaced. A Bayesian CNN has been proposed to recover phase from a single-frequency fringe pattern while providing uncertainty maps that quantify confidence, thereby improving interpretability and reliability assessment \cite{feng2021deep}. However, this uncertainty analysis applies only to single-frequency inference, and the absolute phase estimated from a single fringe remains unreliable. Bayesian CNNs that use a Monte Carlo (MC) method require many stochastic passes, which increases latency and computation and makes uncertainty sensitive to hyperparameters such as dropout rate.

To improve real-time CFPP accuracy while providing uncertainty-aware, interpretable pixel-wise predictions, we propose HSURE-CFPP (Heteroscedastic Snapshot-ensemble Uncertainty-aware Ratio Estimation for CFPP). Here, “ratio” denotes the per-pixel numerator-denominator representation ($N$, $D$) used to compute the wrapped phase. HSURE-CFPP employs a heteroscedastic snapshot-ensemble U-Net (HSU-Net) that jointly predicts mean ($N$, $D$) and the corresponding pixel-wise uncertainty maps. To the best of our knowledge, this is the first demonstration of a snapshot-ensemble CNN with a heteroscedastic likelihood for single-shot phase demodulation from a composite fringe pattern while simultaneously estimating pixel-level uncertainty for both outputs. Experiments show that HSURE-CFPP accurately reconstructs transient scenes and yields uncertainty maps that correlate with reconstruction errors, enabling confidence assessment without ground-truth reference data. Moreover, HSURE-CFPP has a lower inference cost and simpler deployment than Bayesian CNN alternatives.

\section{Result}
In HSURE-CFPP, a composite pattern, termed the hierarchical-frequency composite fringe (HFCF) and designed in the spectral domain, is projected onto the object. The captured fringe patterns are then processed by an HSU-Net trained under a heteroscedastic likelihood and deployed with a snapshot ensemble of checkpoints from different training epochs, enabling real-time reconstruction with per-pixel uncertainty estimates. The HFCF embeds high-, mid-, and low-frequency modulations to mitigate spectral aliasing and enable single-frame TPU, yielding an absolute phase map at the high frequency. This preserves spatial resolution because the high-frequency component provides higher depth sensitivity. The output uncertainties include data uncertainty and model uncertainty. Data uncertainty, also called aleatoric uncertainty, reflects measurement noise and inherent ambiguity in fringe patterns. Model uncertainty, also called epistemic uncertainty, arises from limited training data or generalisation error and tends to be large for out-of-distribution (OOD) or rarely seen surfaces, i.e., samples that differ markedly from the training distribution in material, geometry, or appearance. These two terms are combined to yield the total uncertainty, and phase uncertainty is obtained via first-order error propagation, providing an estimate directly comparable to the phase error.  

Data (aleatoric) uncertainty is encoded in the predicted per-pixel log-variance, whereas model (epistemic) uncertainty is quantified by the variance of predictions across snapshots. Fig.~1 illustrates the workflow of our method. We use $(\hat{\mu},\,\hat{\sigma}^{2})$ for the mean and variance predicted by an individual snapshot, and $(\bar{\mu},\,\bar{\sigma}^{2})$ for the corresponding ensemble-averaged estimates. The HSU-Net is trained three times on the same HFCF input dataset with different ground-truth targets for phase unwrapping, yielding three trained network instances. For each input HFCF, the trained network outputs four quantities: the mean numerator $\bar\mu_N$ and mean denominator $\bar\mu_D$ of the wrapped phase for each modulated frequency, and the corresponding log-variances \(\log \hat{\sigma}_N^{2}\) and \(\log \hat{\sigma}_D^{2}\) that parameterise per-pixel noise. Here $N$ and $D$ index the numerator and denominator of the wrapped phase, respectively. For the \(i\)-th modulated fringe at frequency \(f_{\mathrm{mod},i}\) (\(i=1,2,3\) for high, medium, and low), the wrapped phase is obtained with $\mathrm{atan2}$ applied to the predicted $\bar\mu_N$ and $\bar\mu_D$. The three wrapped phases are then used to determine the fringe order and to unwrap the phase with TPU. Finally, the absolute phase, together with calibration parameters, yields the 3D point cloud by triangulation. The overall reconstruction pipeline is illustrated in the bottom panel of Fig.~1. Uncertainty is quantified in the variance domain. Per-pixel log-variances are exponentiated to obtain data variances, while model variance is estimated from the variance of predictions across the snapshot ensemble. These are summed in the variance domain to form the total predictive variance for $N$ and $D$, which is then propagated to the wrapped phase by first-order error propagation through $\mathrm{atan2}$ at $\bar\mu_N$ and $\bar\mu_D$, producing the variance of the wrapped phase \(\hat{\sigma}_\phi\). Consequently, the uncertainty estimates yielded for each $N$ and $D$ pair enable a more granular and comprehensive assessment of predictive reliability, thereby facilitating the localisation and attribution of potential error sources across different frequencies and network output channels.

\begin{figure}[htbp]
    \centering
    \includegraphics[width=\textwidth]{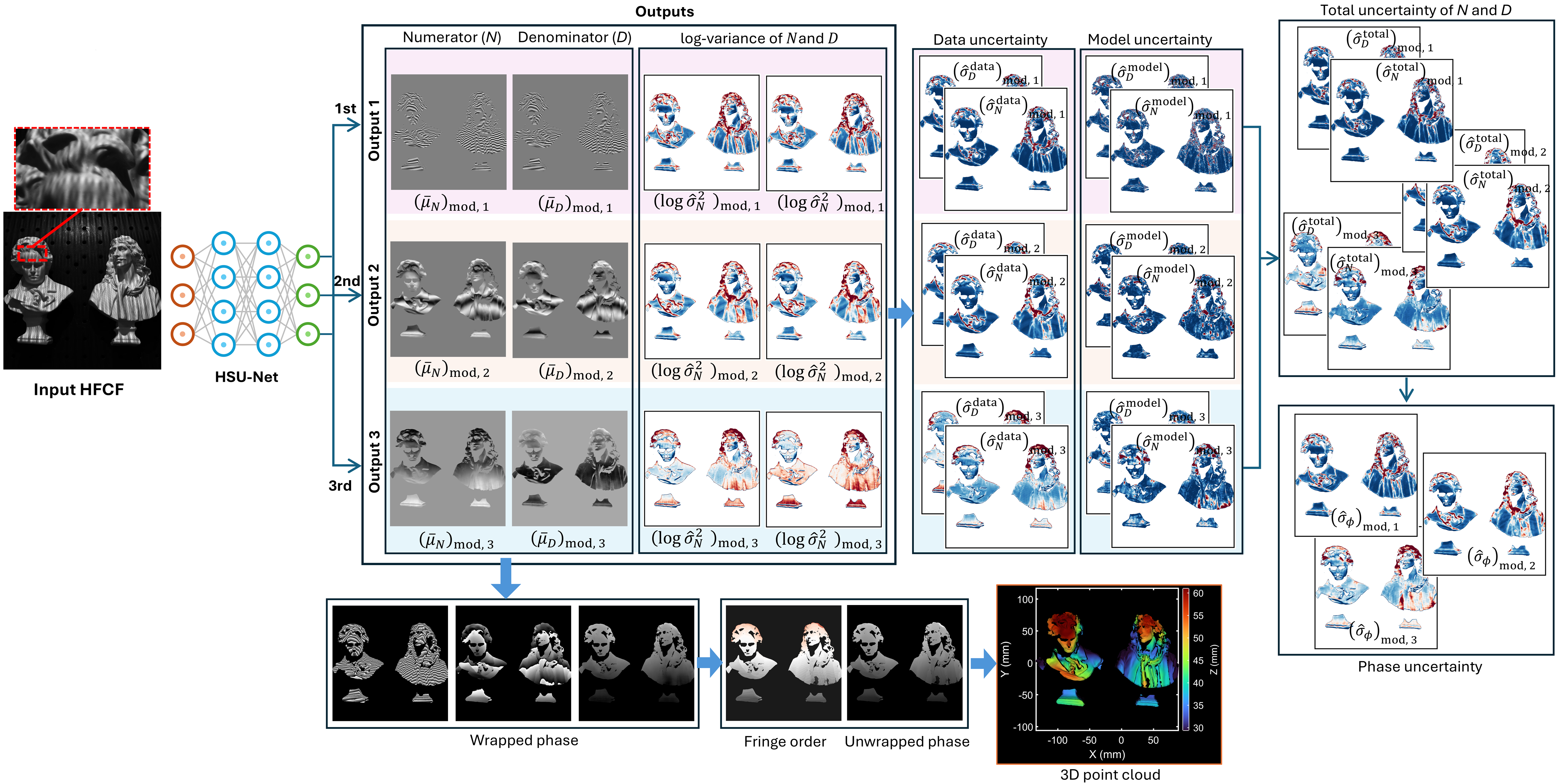}
    \caption{The workflow of the HSURE-CFPP method. An HFCF is fed into the trained HSU-Net to output four quantities, including the mean numerator $\bar\mu_N$ and mean denominator $\bar\mu_D$ and the corresponding log-variances \(\log\hat{\sigma}_N^{2}\) and \(\log \hat{\sigma}_D^{2}\). The means are used to compute wrapped phases, perform phase unwrapping, and reconstruct the 3D point cloud. The log-variances are used to estimate data \(\hat{\sigma}^\mathrm{data}\) and model \(\hat{\sigma}^\mathrm{model}\) uncertainties, which are then combined to obtain the total uncertainty \(\hat{\sigma}^\mathrm{total}\) and the phase uncertainty \(\hat{\sigma}_\phi\) .}
    \label{fig:hsu-workflow}
\end{figure}

\subsection{Principle of HSU-Net CFPP}
\subsubsection{Generation of HFCF}
In HFCF, three modulated fringe patterns with different spatial frequencies whose phase varies only along the vertical direction $y^p$ are encoded into a single high-frequency carrier fringe pattern. Specifically, each vertical modulation term is multiplied by a horizontal carrier with a distinct spatial frequency along $x^p$, and the three products are finally superimposed to form one composite fringe pattern. The $i$-th projected modulated fringe in the projector coordinates $(x^p,y^p)$ is written as

\begin{equation}
\label{eq:projected modulated fringe}
I_{\mathrm{mod},i}^p(x^p,y^p)
= a+ b\,\cos\!\bigl(2\pi f_{\mathrm{mod},i}^p\,y^p\bigr),
\quad i\in\{1,2,3\},
\end{equation}
where $a$ and $b$ are projection constants used to offset $I_{\mathrm{mod}_i}$ to be non-negative values. The $y^p$ axis is the phase direction and corresponds to the vertical direction. The modulation frequencies are hierarchical, $f_{\mathrm{mod},1}>f_{\mathrm{mod},2}>f_{\mathrm{mod},3}$; in our implementation, $f_{\mathrm{mod},3}$ corresponds to one period over the projector resolution. The single-frame HFCF is then given by
\begin{equation}
I_{\mathrm{HFCF}}^p(x^p,y^p)
= \sum_{i=1}^{3}
I_{\mathrm{mod},i}^p(x^p,y^p)\,\cos\!\bigl(2\pi f_{\mathrm{car},i}^p\,x^p\bigr),
\end{equation}
where the carrier frequencies $f_{\mathrm{car}, i}^p$ are oriented along the $x^p$ direction, and ordered from high to low as $f_{\mathrm{car},1}>f_{\mathrm{car},2}>f_{\mathrm{car},3}$. For clarity, Fig.~2(a) illustrates the construction of the HFCF from three modulated fringes and three carrier fringes; the resulting synthetic fringe is shown in Fig.~2(b). Fig.~2(c) displays the 2D Fourier spectrum of the HFCF. $F(f_{\mathrm{mod},0})$ is the zero frequency term and corresponds to the background intensity of the fringe. The three pairs of first-order spectral components, denoted by $F(f_{\mathrm{mod},1}^p)$, $F(f_{\mathrm{mod},2}^p)$, and $F(f_{\mathrm{mod},3}^p)$, arise from the three modulated fringe components and contain the phase information. $f_{\mathrm{car}, i}^p$ and $f_{\mathrm{mod}, i}^p$, are carefully designed so that these first-order lobes are well separated from each other along the horizontal frequency axis (by the different carrier frequencies) and along the vertical frequency axis (by the different modulation frequencies). This prevents spectral aliasing and makes the subsequent network-based learning and feature extraction more stable and reliable. Considering both the intensity capacity of the projector and camera, the intensity of the projected fringe needs to be normalised into $[0,255]$ to map the fringe grayscale to the effective dynamic range of the projection-camera system \cite{zuo2018phase} using 
\begin{equation}
(I_{\mathrm{HFCF}}^p)_{\text{norm}} = \frac{I_{\mathrm{HFCF}}^p - I_{\min}}{I_{\max} - I_{\min}} \times 255
\end{equation}
where $(I_{\mathrm{HFCF}}^p)_{\text{norm}}$ is the normalised composite fringe and will be represented as $I_{\mathrm{HFCF}}^p$ in the rest of the paper, and $I_{\min}$ and $I_{\max}$ are the minimum and maximum intensity values of the original HFCF. 

\begin{figure}[htbp]
    \centering
    \includegraphics[width=\textwidth]{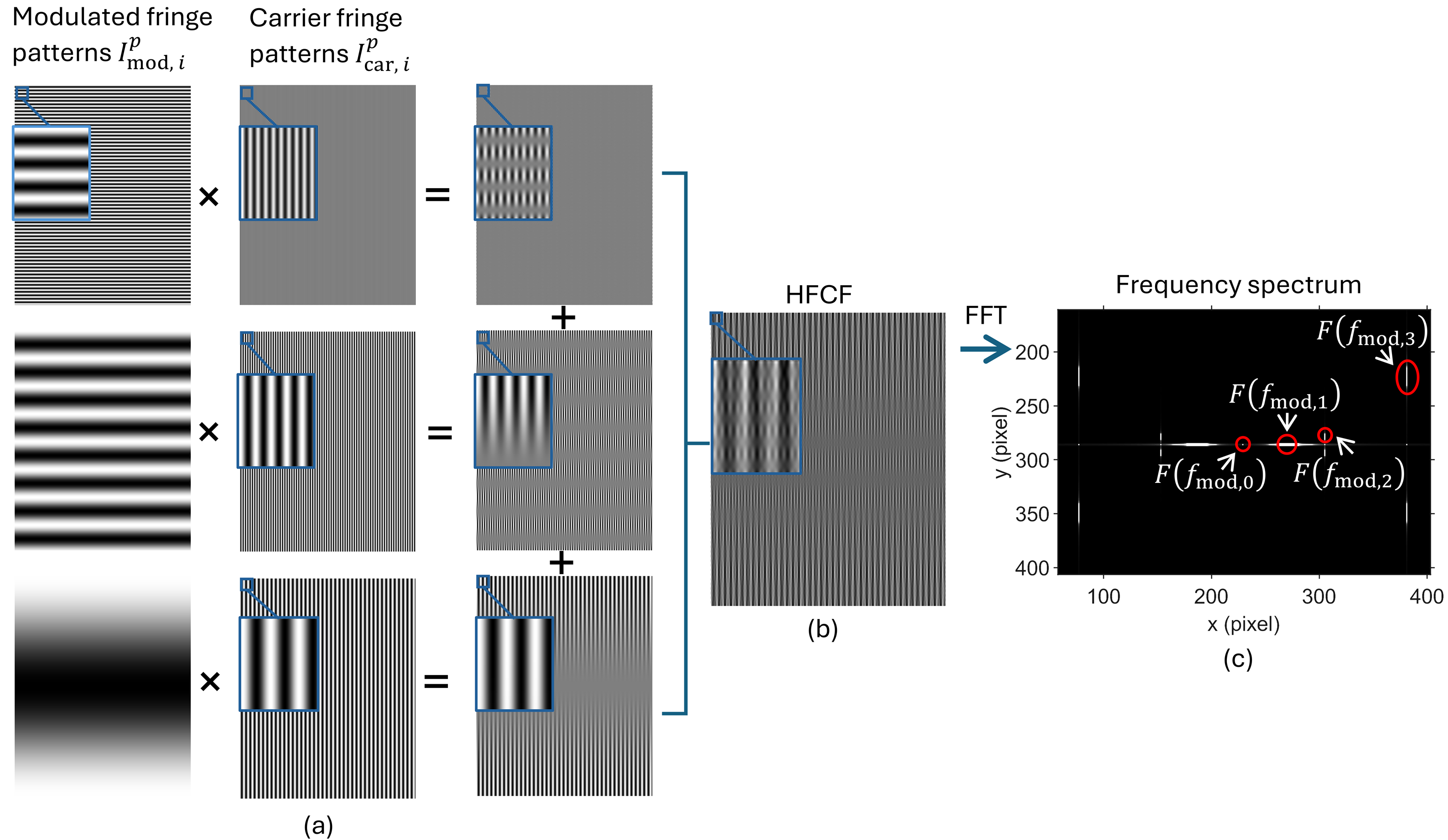}
    \caption{Construction of the HFCF and its Fourier spectrum. (a) Three modulated fringes with different vertical frequencies are multiplied by three carrier fringes with different horizontal frequencies, and the three resulting fringe patterns are then added together to obtain the composite fringe pattern (i.e. HFCF)  in (b). The zoomed-in areas in (a) and (b) show the fringe pattern structure. Three modulated fringes are used to satisfy the TPU criterion for robust phase unwrapping. The size of these fringe patterns is determined by the specifications of the projector. (c) 2D Fourier spectrum of the HFCF, showing the frequency distribution of each modulated component.}
    \label{fig:myfigure}
\end{figure}

After illuminating the measured object with the HFCF using a DLP projector, the captured image can be expressed as
\begin{equation}
    I_{\mathrm{HFCF}}^c(x^c,y^c)
    = R(x^c,y^c) [I_{\mathrm{HFCF}}^p(x^p,y^p) + \beta_1 (x^c,y^c)] + \beta_2 (x^c,y^c)
\end{equation}
where $(x^c, y^c)$ is the pixel coordinates of the camera, $R(x^c,y^c)$ represents the reflectance variation of the measured area. $\beta_1(x^c,y^c)$ models the background illumination on the object surface, while $\beta_2(x^c,y^c)$ accounts for the additive ambient light at the camera and sensor noise. The deformed modulated fringe $I_{\mathrm{mod}, i}^c(x^c, y^c)$ is
\begin{equation}
    \begin{aligned}
I_{\mathrm{mod}, i}^c(x^c, y^c) 
&= R(x^c,y^c) [I_{\mathrm{mod},i}^p(x^p,y^p) + \beta_1 (x^c,y^c)] + \beta_2 (x^c,y^c),\\
&= A(x^c,y^c) + B(x^c,y^c) \cos\!\bigl(\phi(x^c,y^c)\bigr)
    \end{aligned}
\label{eq:captured modulated fringe}
\end{equation}
where $A^p$ and $B^p$ denote the average intensity and modulation amplitude of $I_{\mathrm{mod}, i}^c(x^c, y^c)$, respectively. $\phi(x^c, y^c)$ is the distorted phase that encodes the height information and needs to be calculated. Carrier fringes introduce a spatial frequency shift that separates the modulated components from one another and from the zero frequency in the Fourier domain, enabling HSU-Net to isolate the informative fringe content. The modulation frequencies and the carrier frequencies are designed jointly. The three modulated fringe patterns are used for TPU, and the unwrapped phase at the highest modulation frequency, $f_{\mathrm{mod},1}$, is used for surface reconstruction because of its higher sensitivity to small height variations. Frequency ratios are chosen according to the projector resolution and their locations in the spatial-frequency spectrum, as shown in Fig.~2(c), so that unwrapping remains reliable under noise, saturation, non-uniform reflectance, and occlusion.

\subsubsection{The HSU-CFPP reconstruction principle}
We introduce an HSU-Net (details in Section~4) that directly regresses the sine and cosine components of the wrapped phase from the captured HFCF pattern, which are then used for phase demodulation. According to Eq.~\eqref{eq:captured modulated fringe}, the wrapped phase \(\phi_i\) at the \(i\)-th modulation frequency \(f_{\mathrm{mod},i}^c\) can be recovered from its sine and cosine channels via the four-quadrant arctangent:
\begin{equation}
\label{eq:wrapped_phase_calc}
\phi_i(x^c, y^c)
  = \operatorname{atan2}\!\bigl(N_i(x^c, y^c),\, D_i(x^c, y^c)\bigr),
\end{equation}
where \(N_i(x^c, y^c)\) and \(D_i(x^c, y^c)\) denote the sine and cosine channels of \(\phi_i(x^c, y^c)\), respectively. In the ideal noise-free case, they are proportional to the true sine and cosine of the wrapped phase, i.e., \(N_i(x^c, y^c) \propto \sin\!\bigl(\phi_i(x^c, y^c)\bigr)\) and
\(D_i(x^c, y^c) \propto \cos\!\bigl(\phi_i(x^c, y^c)\bigr)\).

The HSU-Net \(f_{\boldsymbol\theta}\) takes the captured HFCF pattern \(I_{\mathrm{HFCF}}^{c}\) as input and, for each pixel \((x^c,y^c)\), predicts a probabilistic estimate of these two channels. Specifically, the network outputs the mean and log-variance of each channel:
\begin{equation}
\begin{bmatrix}
\bar\mu_{N_i}(x^c,y^c)\\[2pt]
\bar\mu_{D_i}(x^c,y^c)\\[2pt]
\log \hat\sigma_{N_i}^{2}(x^c,y^c)\\[2pt]
\log \hat\sigma_{D_i}^{2}(x^c,y^c)
\end{bmatrix}
= f_{\boldsymbol\theta}\!\big(I_{\mathrm{HFCF}}^{c}\big)\big|_{(x^c,y^c)} ,
\label{eq:hbunet_output}
\end{equation}
where \(\boldsymbol\theta\) denotes the learnable parameters of the network that are optimised on training data. The terms $\mu_{N}$ and $\mu_{D}$ are the predictive means of the numerator and denominator. The terms $\log \hat\sigma_{N}^{2}$ and $\log \hat\sigma_{D}^{2}$ are per-pixel log-variance parameters of a heteroscedastic Gaussian likelihood, which are not involved in reconstruction. The predicted \(\bar\mu_{N_i}\) and \(\bar\mu_{D_i}\) are then used to compute the wrapped phase at each modulation frequency, followed by three-frequency TPU. The highest-frequency wrapped phase \(\phi_3\) is unwrapped with the aid of the lower frequencies. In general, the absolute phase \(\Phi_i(x,y)\) differs from its wrapped counterpart \(\phi_i(x,y)\) by an unknown integer multiple of \(2\pi\):
\begin{equation}
\label{eq:phi_unwrap}
\Phi_i(x,y)=\phi_i(x,y)+2\pi\,k_i(x,y), \qquad k_i(x,y)\in\mathbb{Z}.
\end{equation}
TPU determines the integer map \(k_i\) sequentially from low to high frequency. Assuming the lowest frequency remains unwrapped across the field of view:
\begin{equation}
\Phi_1=\phi_1,\qquad k_1=0.
\end{equation}
For \(i=2,3\), the integer maps \(k_i\)  for the medium and high frequencies are computed as:
\begin{equation}
k_i=\operatorname{round}\!\left(\frac{f_i}{f_{i-1}}\frac{\Phi_{i-1}}{2\pi}-\frac{\phi_i}{2\pi}\right)
\end{equation}
The final absolute phase is \(\Phi_3(x,y)\), which is converted to a 3D point cloud via projector-camera calibration. To train the HSU-Net, we prepare ground-truth targets corresponding to the inputs using a multi-step PSP to obtain high-accuracy \(N_i\) and \(D_i\). Pixel-wise supervision is then applied to encourage the predicted components to match the ground-truth labels in a statistical sense, using a heteroscedastic loss defined over each pixel (see Section~3.5 for the specific loss functions and weighting).

\subsubsection{Design of the Frequencies in HFCF}
Although combining composite fringes with deep learning can improve both speed and accuracy relative to single fringe methods, embedding multiple frequencies in one frame introduces risks. Strong height variations and device nonlinearity can cause spectral crosstalk, high-frequency components are attenuated by the optics, and sampling limits apply. These factors can reduce training stability. Careful design of the carrier and modulation frequencies can mitigate these issues. The carrier frequencies should be mutually well separated to isolate the modulated components in the frequency domain and set as high as practical within system limits to shift these components away from the zero frequency. The modulation frequencies should be chosen according to the phase unwrapping strategy. In this work, a three-frequency TPU scheme is used, so low, medium, and high carriers are selected. Compared with heterodyne approaches that use three closely spaced high frequencies \cite{zuo2016temporal}, this hierarchical frequency design reduces spectral overlap and eases the impact of high-frequency attenuation because only one high-frequency fringe is used. The fringe spatial frequency cannot be arbitrarily increased in pursuit of higher height resolution. The maximum usable spatial frequency is constrained jointly by the camera and projector Nyquist sampling limits, by a minimum-contrast requirement imposed by the system MTF, and by the maximum measurable phase gradient as defined in [20]. These four constraints combine to:

\begin{equation}
\label{eq:kproj-max}
f^{\mathrm{p}}_{\min}
=
\max\!\left\{
2,\;
2\,\frac{\Delta x_{\mathrm{c}}}{\Delta x_{\mathrm{p}}},\;
\frac{1}{\Delta x_{\mathrm{p}}\,f_{\mathrm{MTF}}},\;
\frac{3\,\lvert \nabla \phi \rvert_{\max}}{2\pi}
\right\}.
\end{equation}
where $f^{\mathrm{p}}$ denotes the number of projector pixels per fringe period along the fringe direction, $f_{\min}^{\mathrm{p}}$ is the minimum value ensuring no aliasing and sufficient MTF contrast in theory. $\Delta x^{\mathrm{c}}$ and $\Delta x^{\mathrm{p}}$ are the effective sampling pitch on the object for the camera and the projector, respectively. $f_{\mathrm{MTF}}$ is an MTF-limited frequency ensuring sufficient fringe contrast. $\lvert \nabla \phi \rvert_{\max}$ is the maximum measurable phase gradient (see the supplementary file for more details). By adopting this frequency-selection strategy, we reduce spectral aliasing and preserve contrast, thereby enabling measurement of objects with steep surface gradients, which in turn benefits deep-learning training for multi-frequency demodulation.

\subsubsection{Uncertainty Modeling and Estimation}\label{sec:Uncertainty quantification}
Uncertainty analysis is critical for assessing the reliability of model predictions. BNNs have been used to estimate per-pixel output uncertainty for single-frequency fringes; however, phase ambiguity renders unwrapping brittle, and approximate MC dropout can systematically misestimate uncertainty while incurring substantial inference cost. We address phase ambiguity by using an HFCF that contains multiple phase components at distinct modulation frequencies, thereby increasing the number of independent spectral observations well beyond the unknowns in \eqref{eq:projected modulated fringe}. In practice, sensor noise, spectral aliasing, surface reflectance variations, and fringe attenuation can still lead to unreliable predictions, so the recovered phase may deviate from the ground truth in a data-dependent manner. To trace and analyze such errors, we employ an HSU-Net that retrieves the wrapped phase components \(N\) and \(D\) from an HFCF while simultaneously generating pixel-wise model and data uncertainties through heteroscedastic modeling and a snapshot-ensemble strategy. 

Given a composite fringe $I_{\mathrm{HFCF}}$, HSU-Net outputs for each pixel the predicted $N$ and $D$ together with per-output log-variance parameters that parameterise the heteroscedastic (aleatoric) noise. Concretely, the network produces four per-pixel quantities, $\hat\mu_N,\hat\mu_D,\log\hat{\sigma}_N^{2},\log\hat{\sigma}_D^{2}$. For each pixel $s\in S$, $(\hat\mu_{N,s},\hat\mu_{D,s})$ are treated as point estimates of $N$ and $D$, and $(\hat{\sigma}_{N,s}^{2},\hat{\sigma}_{D,s}^{2})$, with $\hat{\sigma}^2=\exp(\log\hat{\sigma}^{2})$, represent heteroscedastic (aleatoric) noise variances. For $v\in\{N,D\}$, a conditional Gaussian likelihood $v_s^{\ast}\!\sim\!\mathcal{N}(\hat\mu_{v,s},\hat{\sigma}_{v,s}^{2})$ is assumed, and $N$ and $D$ are taken to be conditionally independent given the network weights and the input. The per-pixel negative log-likelihood is
\begin{equation}
\mathcal{L}_{\text{NLL},s}
= \frac{(N_s^{\ast}-\hat\mu_{N,s})^2}{2\,\hat\sigma_{N,s}^2} + \frac{1}{2}\log\hat\sigma_{N,s}^2
+ \frac{(D_s^{\ast}-\hat\mu_{D,s})^2}{2\,\hat\sigma_{D,s}^2} + \frac{1}{2}\log\hat\sigma_{D,s}^2 .
\end{equation}

During training, the per-pixel NLL is computed, averaged over all valid pixels within each image, and then averaged across mini-batches, so the objective reflects image-level rather than isolated-pixel performance. The NLL encourages uncertainty calibration by pushing the predicted per-pixel variance to match the magnitude of the empirical residuals. When the predicted variance is underestimated and large residuals occur, inverse-variance weighting assigns a larger penalty. When the predicted variance is overestimated and inflated, the logarithmic variance term penalises that inflation. As a result, small variances (e.g. $-7\times 10^{-3})$ tend to be produced where predictions are accurate, and larger variances (e.g. $-2\times 10^{-3})$ are produced where the data are intrinsically noisy. A mixed total loss is then optimised that balances mean accuracy and variance learning:
\begin{equation}
\label{eq:total_loss}
\mathcal{L}=\alpha\,\mathcal{L}_{\mathrm{MSE}}+\beta\,\mathcal{L}_{\mathrm{NLL}} .
\end{equation}
here, $\mathcal{L}_{\mathrm{MSE}}$ denotes the mean-squared error over the two outputs $\hat\mu_{N}$ and $\hat\mu_{D}$. It is computed as
\begin{equation}
\mathcal{L}_{\mathrm{MSE}}=\frac{1}{|S|}\sum_{s\in S}\Big[(\hat\mu_{N,s}-N_s^{\ast})^2+(\hat\mu_{D,s}-D_s^{\ast})^2\Big] .
\end{equation}
The coefficient $\beta$ is warmed up from $0$ to its target value during the early epochs (we fix $\alpha=1.0$ and $\beta=1.0$, and linearly increase $\beta$ from 0 to 0.04 over the first 30 epochs), so that when $\beta$ is small the optimisation emphasises $\mathcal{L}_{\mathrm{MSE}}$ to stabilise the means $(\hat\mu_N,\hat\mu_D)$, and as $\beta$ increases $\mathcal{L}_{\mathrm{NLL}}$ aligns the predicted variances with the residual magnitude and discourages trivial variance inflation. Epistemic uncertainty is quantified without learning a weight posterior by using a snapshot ensemble. In contrast to BNNs, no posterior over weights or per-layer dropout rates is learned, and dropout is used solely as a standard regulariser and is disabled at test time. Cosine annealing with warm restarts produces $T$ deterministic checkpoints $\{\boldsymbol{\theta}^{(t)}\}_{t=1}^{T}$. For a test image $x^{\ast}$, checkpoint $t$ yields $\hat\mu_{v,s}^{(t)}=\mu_v(x^{\ast},s;\boldsymbol{\theta}^{(t)})$ and $\hat\sigma_{v,s}^{2,(t)}=\sigma_v^2(x^{\ast},s;\boldsymbol{\theta}^{(t)})$. The ensemble mean is:

\begin{equation}
\label{eq:ensemble_mean}
\bar\mu_{v,s}=\frac{1}{T}\sum_{t=1}^{T}\hat\mu_{v,s}^{(t)}.
\end{equation}

The data uncertainty is obtained by averaging the heteroscedastic variances predicted by each checkpoint:
\begin{equation}
\label{eq:data_uncertainty}
\bar{\sigma}_{v,s}^{2,\text{data}}
= \frac{1}{T}\sum_{t=1}^{T}\hat{\sigma}_{v,s}^{2,(t)} .
\end{equation}

The model uncertainty is measured by the dispersion of snapshot means around the ensemble mean:
\begin{equation}
\label{eq:model_uncertainty}
\bar{\sigma}_{v,s}^{2,\text{model}}
= \frac{1}{T}\sum_{t=1}^{T}\big(\hat{\mu}_{v,s}^{(t)}-{\bar{\mu}}_{v,s}\big)^{2}.
\end{equation}

The total predictive variance is obtained by summing the data and model uncertainty terms:
\begin{equation}
\label{eq:total_uncertainty}
\bar{\sigma}_{v,s}^{2,\text{total}}
= \bar{\sigma}_{v,s}^{2,\text{data}}+\bar{\sigma}_{v,s}^{2,\text{model}} .
\end{equation}

Here, $\bar{\sigma}^{2,\text{data}}_{v,s}$ captures inherent data noise predicted by the network, whereas $\bar{\sigma}^{2,\text{model}}_{v,s}$ measures dispersion across checkpoints and typically decreases with improved data coverage or model capacity. $\bar{\sigma}^{2,\text{total}}_{v,s}$ aggregates these two sources into a single reliability measure. It increases when either component is large and is the quantity we propagate to the phase and visualise as confidence maps. To propagate the uncertainty of $N$ and $D$ to the wrapped phase $\phi$, first-order error propagation is carried out around the ensemble mean estimates. Let $N_s=\bar{\mu}_{N,s}$ and $D_s=\bar{\mu}_{D,s}$, the Jacobian at $(N_s,D_s)$ is

\begin{equation}
\label{eq:Jacobian}
\frac{\partial \phi_s}{\partial N_s}=\frac{D_s}{N_s^2+D_s^2}, \qquad
\frac{\partial \phi_s}{\partial D_s}=-\,\frac{N_s}{N_s^2+D_s^2}.
\end{equation}

Assuming independent perturbations of $N_s$ and $D_s$, the phase variance is given by:
\begin{equation}
\label{eq:phase_uncertainty}
\begin{aligned}
\bar{\sigma}_{\phi,s}
&= \left(\frac{\partial \phi_s}{\partial N_s}\right)^{2}\bar{\sigma}_{N,s}^{2,\text{total}}
 + \left(\frac{\partial \phi_s}{\partial D_s}\right)^{2}\bar{\sigma}_{D,s}^{2,\text{total}} \\[4pt]
&= \frac{ D_s^{2}\,\bar{\sigma}_{N,s}^{2,\text{total}} + N_s^{2}\,\bar{\sigma}_{D,s}^{2,\text{total}} }
        { (N_s^{2}+D_s^{2}+\varepsilon)^{2} } .
\end{aligned}
\end{equation}

In practice, we add a tiny $\varepsilon$ (e.g. $-8\times 10^{-3})$ to the denominator for numerical safety and mask pixels where $N_s^{2}+D_s^{2}$ is extremely small. For visualisation, standard deviations are reported as the square roots of the corresponding variances. All combinations and propagation are carried out in the variance domain. The probabilistic grounding of aleatoric (data) uncertainty is preserved, and model averaging for epistemic (model) uncertainty is approximated by a snapshot ensemble of deterministic checkpoints. Phase-domain confidence maps ($\hat\sigma_\phi$) are produced and are suitable for quality assessment and reconstruction weighting. Fig.~3 illustrates how the outputs of HSU-Net are used for phase unwrapping and uncertainty estimation.

\begin{figure}[htbp]
    \centering
    \includegraphics[width=\textwidth]{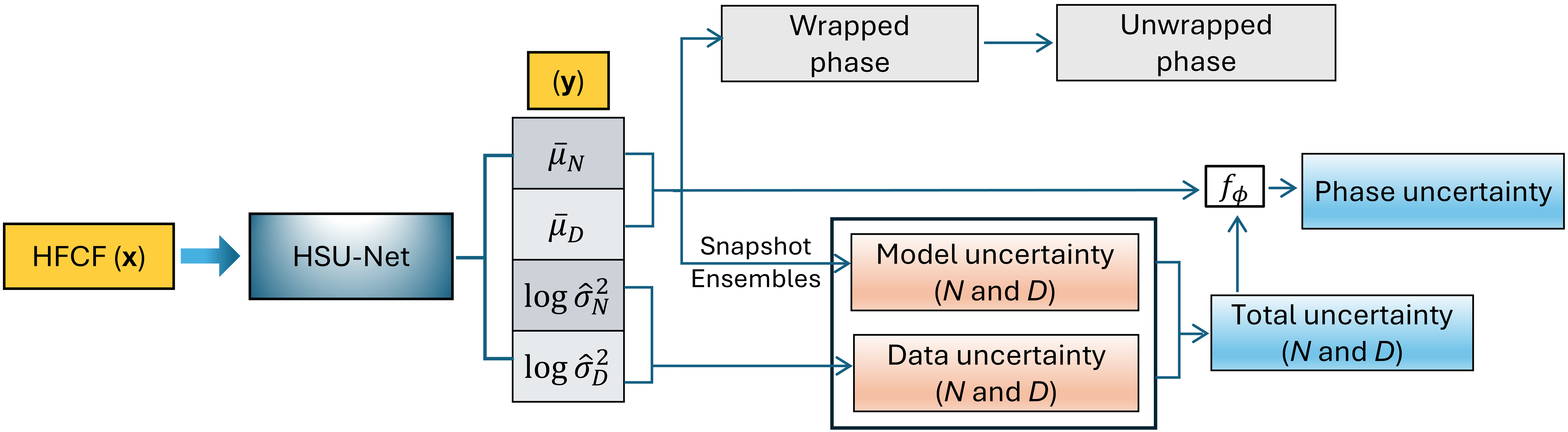}
    \caption{Uncertainty estimation pipeline using HSU-Net predictions. The HFCF is fed into HSU-Net to predict the means \(\bar\mu_{N}\) and \(\bar\mu_{D}\) and the corresponding log-variances, which are used to compute the data uncertainty. A snapshot ensemble provides multiple estimates of \(\hat\mu_{N}\) and \(\hat\mu_{D}\), from which the model uncertainty is computed. The data and model uncertainties are then combined to obtain the total uncertainty and the phase uncertainty.}
    \label{fig:myfigure}
\end{figure}

\subsection{HSU-Net architecture}
The idea of HSU-Net is inspired by the recent successful applications of deep learning techniques on FPP, such as CNNs \cite{zuo2022deep}. Considering that the objective of HSU-Net is to accomplish a pixel-wise image-to-image transformation task, image segmentation networks, such as CNN-based U-Net, ResNet, and U-Net derivative networks, are preferred options \cite{falk2019u,siddique2020u}. Informed by prior studies, we adopt a standard encoder–decoder U‑Net as the backbone by taking into account its functionality and practicality. For the baseline U-Net as shown in Fig.~4, the input is a single–channel grayscale HFCF. The encoder contains four convolutional blocks, each followed by a \(2\times2\) max–pooling layer with a stride of \(2\). The encoder channel widths are \(\{1,\,64,\,128,\,256,\,512\}\). Every block consists of two consecutive \(3\times3\) convolutions (stride \(1\), padding \(1\)) followed by batch normalisation and ReLU activations. A Dropout2d layer with probability \(p=0.1\) is inserted between the two convolutions to mitigate overfitting. The bottleneck employs the same block structure with 1024 channels. The decoder mirrors the encoder. Feature maps are upsampled using \(2\times2\) transposed convolutions with stride \(2\) and are concatenated with the correspondingly sized encoder features through skip connections. Each concatenation is followed by a convolutional block identical to those in the encoder. The decoder channel widths are \(\{1024,\,512,\,256,\,128,\,64\}\). Finally, a \(1\times1\) convolution projects the features to the required number of output channels. This architecture preserves high–resolution spatial details via skip connections while exploiting the large receptive field available in deeper layers for global consistency \cite{siddique2020u}. Unless otherwise stated, we use zero padding to keep the feature-map size unchanged within each convolutional block. 

Building on a baseline encoder--decoder U-Net, we append a heteroscedastic regression head to jointly predict pixel-wise means and log-variances for $(N,D)$. At inference, we employ a snapshot ensemble with $T=4$ deterministic checkpoints $\{\boldsymbol{\theta}^{(t)}\}_{t=1}^{T}$; checkpoint $t$ produces $(\hat{\mu}_{N}^{(t)},\hat{\mu}_{D}^{(t)})$ and $(\log \hat{\sigma}_{N}^{2,(t)}, \log \hat{\sigma}_{D}^{2,(t)})$, and the final mean estimates $(\bar{\mu}_{N},\bar{\mu}_{D})$ are obtained by averaging the checkpoint means. The checkpoint-wise means and variances are further combined to form uncertainty maps as described in Section~\ref{sec:Uncertainty quantification}. Training minimises a mixed objective given by a weighted sum of $\mathcal{L}_{\mathrm{MSE}}$ and $\mathcal{L}_{\mathrm{NLL}}$ with weights $\alpha$ and $\beta$, respectively, which preserves HSU-Net point-estimation performance while enabling heteroscedastic variance learning and better-calibrated uncertainty under spatially varying noise. For phase unwrapping, we train three HSU-Net models with the same HFCF inputs but frequency-specific ground-truth targets (high-, medium-, and low-frequency), yielding three sets of learned network parameters $\{\boldsymbol{\theta}_{\text{high}}, \boldsymbol{\theta}_{\text{mid}}, \boldsymbol{\theta}_{\text{low}}\}$ for predicting $(\hat{\mu}_{N},\hat{\mu}_{D})$ at each modulation frequency.

\begin{figure}[htbp]
    \centering
    \includegraphics[width=\textwidth]{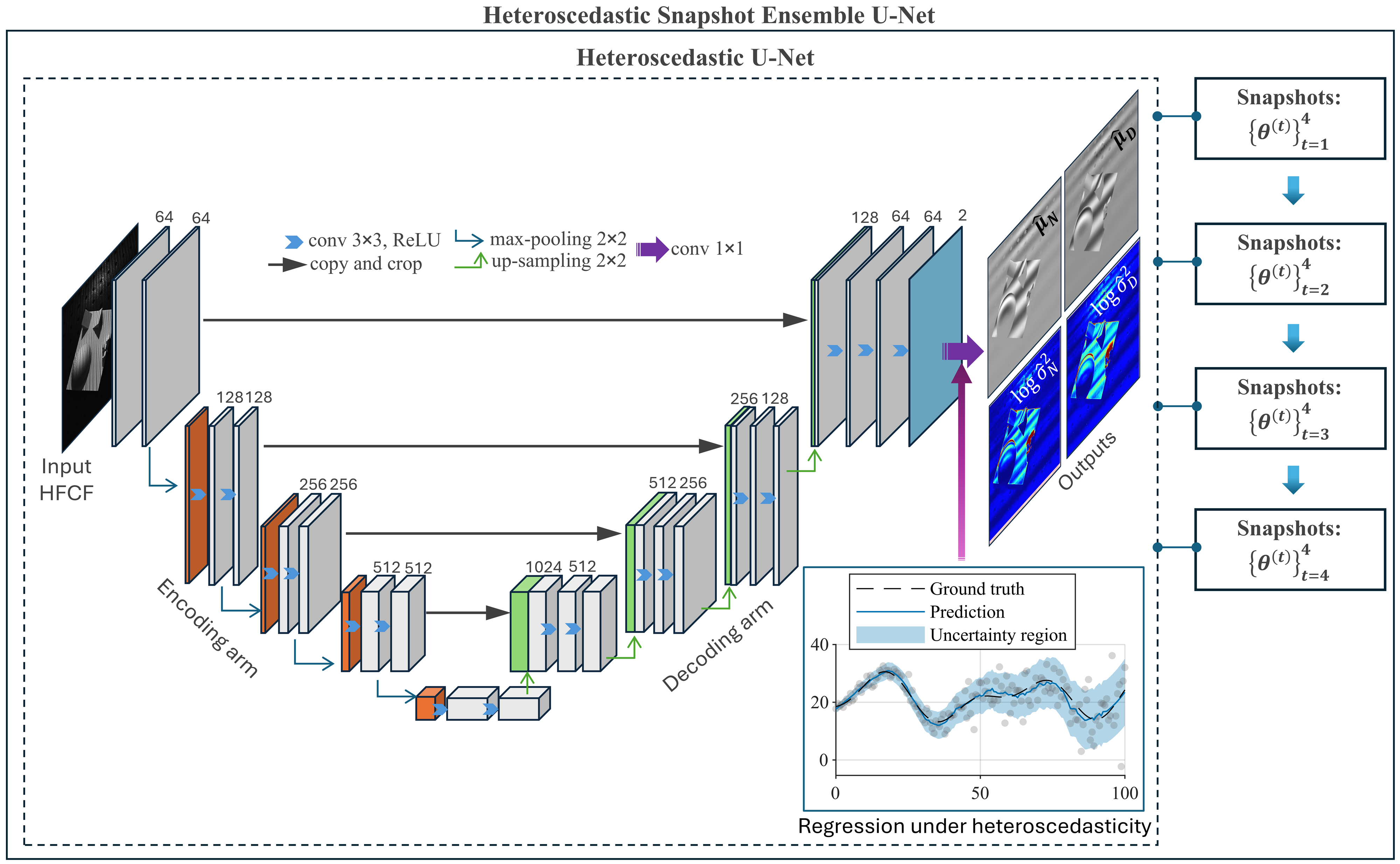}
    \caption{Our HSU-Net architecture builds on a U-Net backbone that takes the HFCF input and, via an encoder--decoder with skip connections, uses a heteroscedastic regression head to jointly predict $(\hat{\mu}_N,\hat{\mu}_D)$ and $(\log\hat{\sigma}_N^2,\log\hat{\sigma}_D^2)$, enabling pixel-wise aleatoric uncertainty estimation alongside $(N, D)$ regression; a snapshot ensemble aggregates multiple checkpoints to improve robustness and quantify epistemic uncertainty.}
    \label{fig:myfigure}
\end{figure}

\subsection{Dataset construction}\label{sec:dataset_construction}
In order to feed the dataset to HSU-Net for network training, we construct a high-quality dataset including the input data and the ground truth. The input data comprises single-frame fringe patterns captured from various objects, while the corresponding ground truth data are the \(N_i\) and \(D_i\), which are calculated from multi-frame fringe patterns of the same objects using a traditional multi-step PSP approach with the same measurement setup. These multi-step PSP facilitate the calculation of high-quality numerators and denominators of the wrapped phase, as well as the derivation of absolute phase and 3D point clouds, which serve as a benchmark for validating the predictions of our model.

In PSP, for the \(i\)-th fringe frequency, the intensity of the \(m\)-th phase-shifted deformed fringe \(I_{m,i}(x^c,y^c)\) at pixel \((x^c,y^c)\) is modeled as
\begin{equation}
I_{m,i}(x^c,y^c)
= R(x^c,y^c)(A(x^c,y^c) + B(x^c,y^c)\cos\!\left[\phi_m(x^c,y^c) - \frac{2\pi m}{M-1}\right]).
\label{eq:fringe_model}
\end{equation}
where \(\phi_i(x^c,y^c)\) represents the wrapped phase at the \(i\)-th fringe frequency $f_i$ with \(i\in\{1,2,3\}\). The subscript \(m\) indexes the phase-shifted fringe with \(m\in\{0,1,2,\ldots,M-1\}\). In our framework, the fringe frequencies used in the PSP procedure for dataset generation are chosen to be identical to the spatial modulation frequencies \(f_{\mathrm{mod},i}\) of the HFCF fringes that the network is trained to recover. Matching these frequency sets ensures that the supervisory signals correspond to the same underlying phase field as the input HFCF patterns. Moreover, once phase wrapping is involved, changing the fringe frequency alters the sampling of the phase and the locations of the \(2\pi\) discontinuities. Consequently, labels computed at one frequency cannot, in general, be converted to labels at another frequency by a simple linear scaling of the phase. For the \(i\)-th fringe frequency, the wrapped phase at pixel \((x^c,y^c)\) can be computed from an \(M\)-step phase-shifting fringes as:
\begin{equation}
\begin{aligned}
N_{i}^\mathrm{gt}(x^c,y^c)
  &= \sum_{m=0}^{M-1} I_{m,i}(x^c,y^c)\,
       \sin\!\Bigl(\frac{2\pi m}{M-1}\Bigr), \\
D_{i}^\mathrm{gt}(x^c,y^c)
  &= \sum_{m=0}^{M-1} I_{m,i}(x^c,y^c)\,
       \cos\!\Bigl(\frac{2\pi m}{M-1}\Bigr),
\end{aligned}
\label{eq:ND_gt}
\end{equation}
Based on these two components, the wrapped phase at pixel \((x^c,y^c)\) is obtained as
\begin{equation}
\phi_{i}(x^c,y^c)
  = \operatorname{atan2}\!\bigl(N_{i}^\mathrm{gt}(x^c,y^c),\,
                                 D_{i}^\mathrm{gt}(x^c,y^c)\bigr).
\label{eq:wrapped_phase_m}
\end{equation}
where \(N_{i}^{\mathrm{gt}}(x^{c},y^{c})\) and \(D_{i}^{\mathrm{gt}}(x^{c},y^{c})\) are used as the ground-truth labels for the cosine and sine channels of the wrapped phase, respectively. The absolute phase is then computed via TPU as described in Section~\ref{sec:dataset_construction}. The large phase-shift number \(M\) can effectively suppress various types of phase errors, such as phase-shifting error and system nonlinearity errors. To reduce phase-calculation error while maintaining measurement efficiency, we use PSP with different numbers of phase shifts at the three modulation frequencies: \(M_{1}=9\), \(M_{2}=6\), and \(M_{3}=6\) for the high-, medium-, and low-frequency fringes, respectively. Since the absolute phase for 3D reconstruction is derived from the high-frequency fringe, a larger number of phase shifts is used at this frequency than at the medium and low frequencies to enhance the robustness of the wrapped-phase estimation. In total, the measurement system will project and synchronously capture 21 fringe patterns and one single HFCF on each target object for constructing the training dataset. Massive and diverse datasets are frequently required by the supervised deep learning networks to ensure reliable training. Therefore, the measured objects included in the dataset with various sizes and geometries are randomly positioned and oriented and captured by a camera to increase the number and diversity of the dataset. Moreover, to increase the generalisation ability of the model, several single objects are randomly grouped and placed together, connected or isolated, to supplement the dataset. Totally, the generated dataset contains 1000 data samples and is split into a training dataset, validation dataset, and test dataset by a ratio of 8:1:1. The training and validation datasets are used to determine the network weight and stop training when the model converges. The test dataset is used to evaluate the performance of the trained model that never appears in other datasets. The dataset augmentation technique, such as image rotation and image flipping, is not implemented in our dataset. Because those operations violate the geometric constraints of the calibration and are equivalent to changing the position of the camera and projector, this leads to unstable training and reduced model generalisation. Only the cropping operation is applied in captured images for locating the region of interest (ROI) to increase the training efficiency.

\subsection{Experiments}
A comprehensive quantitative experiment was performed to verify the effectiveness and robustness of our proposed HSU-Net CFPP approach for dynamic 3D shape measurement and reconstruction. After hyperparameter tuning of the HSU-Net model, the proposed method is tested and analysed with different object samples in the test dataset. The proposed method is evaluated by comparing the predicted results with the ground truth, using quantitative error metrics on the wrapped phase and the reconstructed height map. Then, the uncertainty maps that describe the model uncertainty and data uncertainty are quantified. Finally, the conclusion related to the HSU-Net-CFPP is provided based on the experimental analysis.

\subsubsection{Experimental system}
The measurement system consists of a DLP projector (LightCrafter 4500) with a resolution of 912 $\times$ 1140 and a monocular machine vision camera (Basler acA5472-17uc) with a resolution of 5472 $\times$ 3648 with a 16\,mm focal length lens (Basler C23-1616-2M-S). The maximum frame rate of the camera is 17 fps, and the projector is 120\,fps when 8-bit grey patterns are projected. With the aperture set to f/11 (lens minimum aperture: f/16) and an exposure time of 50 ms, the frame period was set to 110 ms, with the camera shutter mode was Global Reset Release. Our measurement system provides a measurement field of view of approximately $8.73\,\mathrm{cm}^2$ with a spatial resolution of about $50\,\mu\mathrm{m}$. To meet GPU memory constraints, each raw frame ($5472 \times 3648$ pixels) is center-cropped to a $3648 \times 3648$-pixel square ROI by removing 912 pixels from both the left and right margins. The ROI is then downsampled to $912 \times 912$ pixels using OpenCV's area-based resampling (INTER\_AREA), which preserves fine details while effectively suppressing aliasing and moiré artifacts. After preprocessing, the ROI corresponds to an imaged area of approximately $5.29\,\mathrm{cm}^2$ and yields a compact network input that reduces computational cost. The computer components used in the experiments include a Lenovo ThinkStation P3 Tower equipped with an Intel® Core™ i9-13900K processor, 128 GB of RAM, and an NVIDIA GeForce RTX 4070 GPU. The network is implemented and trained in PyTorch 2.5.1 (Python 3.9.18) with GPU acceleration, using the official NVIDIA CUDA build (cu118). 

\subsubsection{HSU-Net Implementation Details}
Hyperparameters were tuned for stable convergence, notably the learning rate, number of epochs, batch size, and number of snapshots. Training minimises a composite objective combining MSE (weight \(1.0\)) and NLL (weight \(0.04\)); the NLL contribution is linearly warmed up over the first \(30\) epochs to stabilise variance learning. Optimisation uses AdamW (initial learning rate \(5\times 10^{-4}\), weight decay \(1\times 10^{-4}\)). The learning rate follows cosine annealing with warm restarts (20-epoch cycle), and snapshots are saved at epochs \(20, 40, 60, 80\) to form an ensemble for model-uncertainty estimation. Each convolutional block applies ReLU activations with \(0.1\) dropout between the two convolutions to curb overfitting. Training runs for up to \(80\) epochs with a mini-batch size of \(4\); early stopping halts optimisation after \(25\) epochs without validation improvement. To enhance stability and throughput, gradients are clipped to an \(\ell_{2}\)-norm of \(1.0\) and automatic mixed precision (float16) with dynamic loss scaling is enabled. This configuration accelerates convergence, improves robustness to heteroscedastic pixel-level noise, and yields well-calibrated predictive uncertainties. Fig.~6 plots the training and validation losses for the three frequency-specific models trained to regress \((N_i,D_i)\) at high, medium, and low modulation frequencies. In all cases, both the total loss (MSE\,+\,NLL) and the MSE term decrease over epochs and approach a plateau, indicating convergence. The total loss can become negative because the heteroscedastic Gaussian NLL is evaluated without its constant normalisation term. When the residuals are small and the model predicts low variances, the NLL term may take negative values even though the MSE component remains positive. Training three frequency-specific models on the same inputs but with different targets (high-, medium-, and low-frequency \((N_i,D_i)\)) takes a totals 42 hours. Inference requires 1.94\,s per image to generate \((N_i, D_i)\), and the corresponding uncertainty maps. For time-critical measurements, the captured fringes can be buffered during acquisition and processed offline. 

\begin{figure}[htbp]
    \centering
    \includegraphics[width=\textwidth]{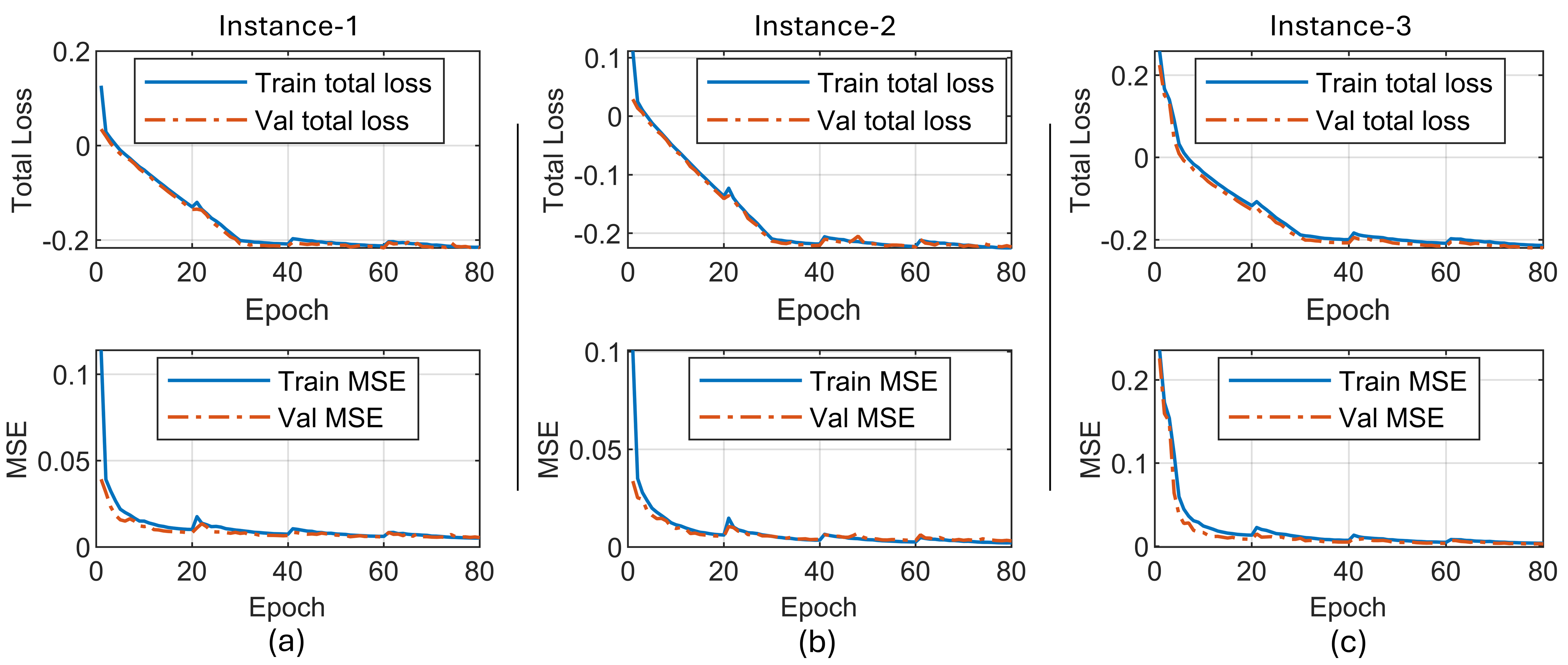}
    \caption{Training and validation curves for the total loss (weighted sum of MSE and NLL) and for MSE show stable convergence across the three frequency-specific models. Fig. (a)–(c) depict three model instances trained on the same HFCF input but different ground-truth targets \((N_i^{\mathrm{gt}}, D_i^{\mathrm{gt}})\) at high-, medium-, and low-modulation frequencies. Both metrics decrease and plateau over epochs. Because the Gaussian NLL excludes the constant normalisation term, the total loss can be negative.}
    \label{fig:myfigure_1}
\end{figure}

\subsubsection{Quantitative Analysis}
The proposed HSU-Net-based CFPP method recovers the sine and cosine components \(N\) and \(D\) of the wrapped phase from a single HFCF pattern, which are then used for phase unwrapping and real-time 3D reconstruction. Since the 3D shape is  derived from the unwrapped phase, the prediction accuracy of \(N\) and \(D\) largely determines the final reconstruction accuracy. To evaluate the proposed method, we conduct experiments on test objects and scenes that were not seen during training and compare the reconstructed results with the corresponding ground-truth data. As shown in Fig.~6, two HFCFs, one corresponding to a single object and the other to multiple isolated objects, are fed into the trained model to infer \(N\) and \(D\) together with their log-variances, in order to assess out-of-distribution generalisation. Since each HFCF contains three modulated fringes, the model outputs three pairs \(\bigl((\mu_N)_{\mathrm{mod},i},\, (\mu_D)_{\mathrm{mod},i}\bigr)\) at distinct modulation frequencies, together with the corresponding log-variance parameters \(\bigl((\log \sigma_N^2)_{\mathrm{mod},i},\, (\log \sigma_D^2)_{\mathrm{mod},i}\bigr)\). The pairs \((\mu_N)_{\mathrm{mod},i}\) and \((\mu_D)_{\mathrm{mod},i}\) are used to compute the wrapped phase and fringe order for phase unwrapping, while the log-variance parameters are used to construct uncertainty maps. In the fringe analysis, including wrapped phase computation and fringe order determination, we apply the composite mask shown in Fig.~6 to remove outliers. This composite mask is defined as the intersection of an amplitude mask obtained from the phase-shifted fringes and a height mask specified by the valid height range of the object.

\begin{figure}[htbp]
    \centering
    \includegraphics[width=\textwidth]{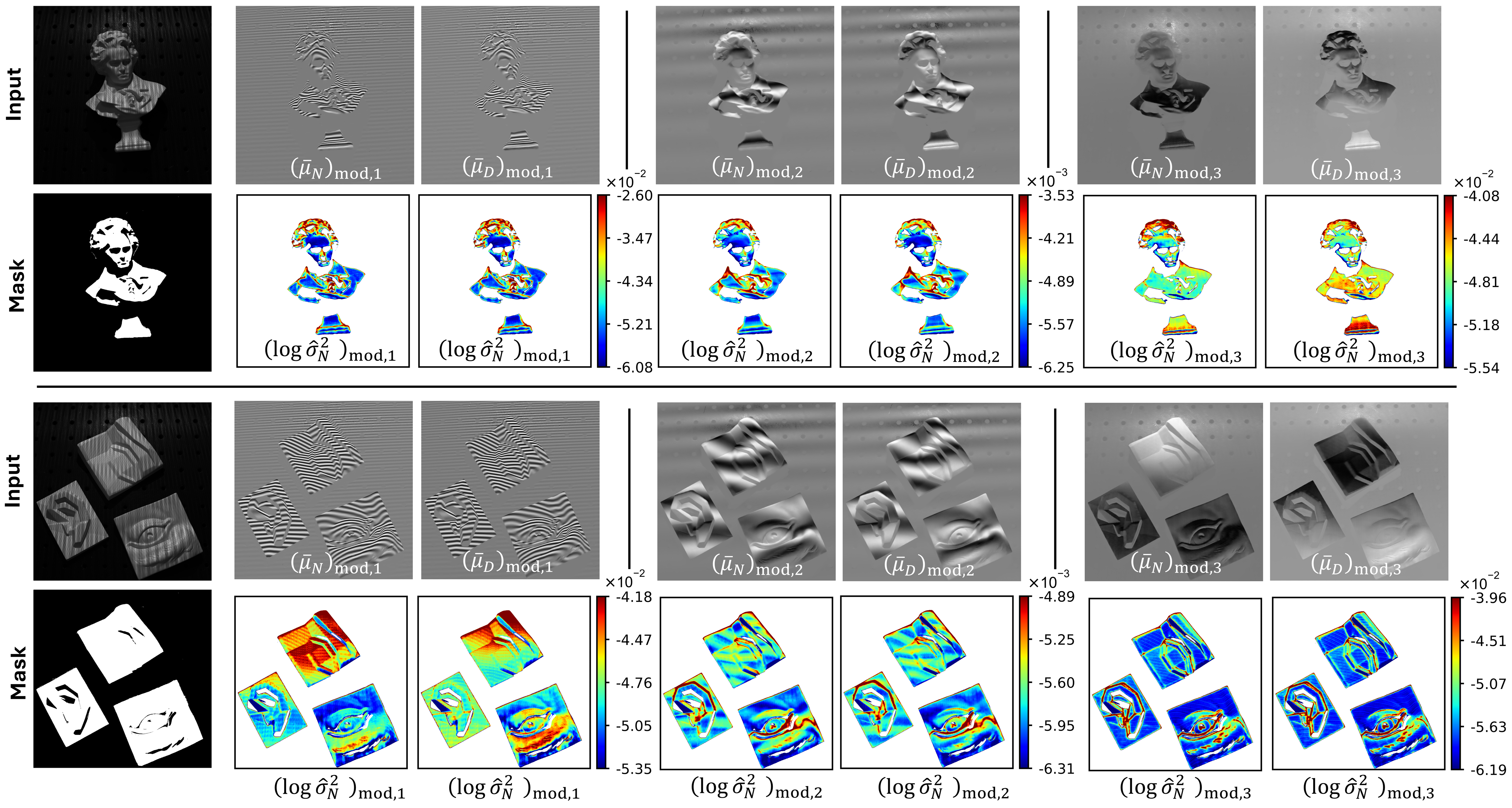}
    \caption{Two object scenarios of different types are measured to evaluate HSU-Net’s reliability and generalisation. The HFCFs from both scenarios are used as inputs to the trained model; inference yields the means of $N$ and $D$ at three modulation frequencies along with their log variances. A composite mask defined by fringe amplitude and the object’s height range is applied for outlier removal.}
    \label{fig}
\end{figure}

To further highlight the strengths of our approach, we also apply the FTP method to the same object, and the resulting point cloud is shown in Fig.~7(a). For a fair comparison with HSU-Net, we train an in-house enhanced U-Net baseline with batch normalization and dropout, and enable automatic mixed precision (AMP) with gradient scaling (GradScaler) \cite{garbin2020dropout}. We use the same dataset, hyperparameters, and encoder--decoder channel widths as HSU-Net. The loss function is $\mathcal{L} = \mathrm{MSE}(\mu_{N}, N) + \mathrm{MSE}(\mu_{D}, D)$. The reconstructed 3D point cloud using the predictions from the enhanced U-Net is shown in Fig.~7(b), the 3D point cloud obtained with our HSU-Net appears in Fig.~7(c), and the ground-truth point cloud in Fig.~7(d) is obtained using a nine-step PSP method. Across the four point clouds, FTP yields noticeably lower quality than the ground truth; fine surface details are missing, and ripple-like artefacts appear, which we attribute to high-frequency spectral loss and spectral aliasing. The enhanced U-Net provides an alternative for phase unwrapping with our HFCF input, but does not provide uncertainty estimates. Its accuracy remains inferior to HSU-Net because it assumes homoscedastic noise and lacks per-pixel uncertainty modelling, which reduces robustness in low-SNR and high-frequency regions. For the two object scenarios in Fig.~6, the height MAE values are \(1.37\,\mathrm{mm}\) and \(0.95\,\mathrm{mm}\) for FTP, \(0.22\,\mathrm{mm}\) and \(0.11\,\mathrm{mm}\) for the enhanced U-Net, and \(0.12\,\mathrm{mm}\) and \(0.03\,\mathrm{mm}\) for HSU-Net. The scenario with multiple isolated objects attains a lower error than the single-object scenario. We attribute this to the more complex geometry present in the single object, whereas the isolated objects have smoother surfaces. This suggests that surface complexity is a primary factor that limits reconstruction accuracy, a trend that is corroborated by the subsequent uncertainty analysis.

\begin{figure}[htbp]
    \centering
    \includegraphics[width=\textwidth]{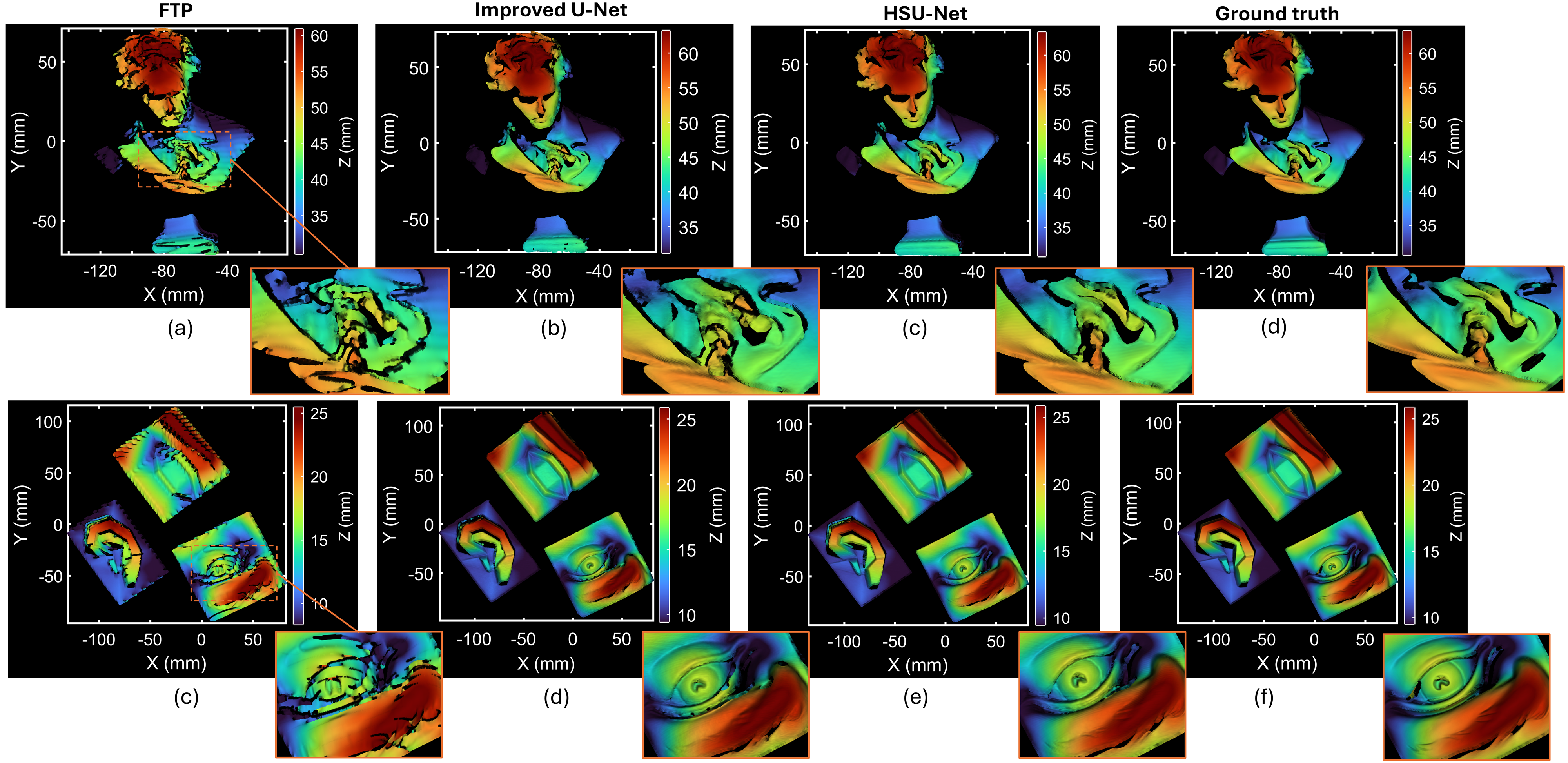}
    \caption{3D point clouds of the two objects shown in Fig.~6 using different methods, including FTP (a), an improved U-Net (b), and our method (c), with comparison to the ground-truth point cloud (d) obtained by the conventional nine-step PSP. Among the methods, FTP yields the lowest accuracy, and the improved U-Net performs slightly worse than our method when evaluated against the ground truth.}
    \label{fig}
\end{figure}

\subsubsection{Uncertainty Estimation}
Uncertainty maps provide a calibrated per-pixel reliability representation that indicates where errors are likely and clarifies their sources, thereby improving robustness and accuracy in 3D reconstruction. To compute the uncertainties, we use the model outputs \(\bar{\mu}_{N}\) and \(\bar{\mu}_{D}\) together with their corresponding log-variances, as shown in Fig.~6. Following Eqs.~\eqref{eq:ensemble_mean}--\eqref{eq:model_uncertainty}, we obtain the data uncertainties \((\hat\sigma^{\mathrm{data}}_N)_{\mathrm{mod},i}\) and \((\hat\sigma^{\mathrm{data}}_D)_{\mathrm{mod},i}\) and the model uncertainties \((\hat\sigma^{\mathrm{model}}_N)_{\mathrm{mod},i}\) and \((\hat\sigma^{\mathrm{data}}_D)_{\mathrm{mod},i}\), which are shown in Fig.~8(b). These are then used to compute the total uncertainties \((\hat\sigma^{\mathrm{total}}_N)_{\mathrm{mod},i}\) and \((\hat\sigma^{\mathrm{total}}_D)_{\mathrm{mod},i}\) (Fig.~8(a)) and the phase uncertainty \((\hat\sigma_\phi)_{\mathrm{mod},i}\) (Fig.~8(c)) according to Eqs.~\eqref{eq:total_uncertainty}--\eqref{eq:phase_uncertainty}. 

\begin{figure}[htbp]
    \centering
    \includegraphics[width=\textwidth]{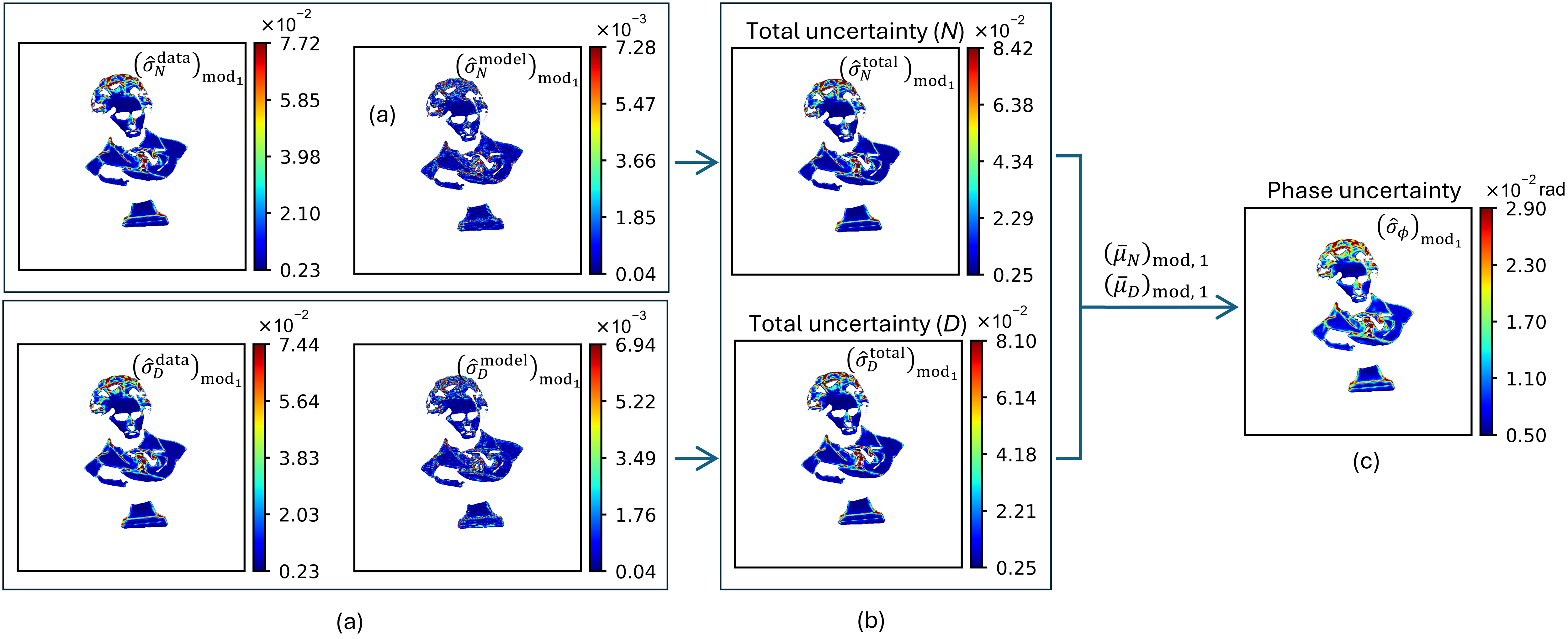}
    \caption{Visualisation of uncertainty maps from HSU-Net. The data and model components are combined to obtain the total uncertainty for \(N\) and \(D\) and are further propagated to yield the phase uncertainty. (a) Data and model uncertainty maps for \(N\) and \(D\). (b) Total uncertainty maps are obtained by combining data and model uncertainties. (c) Phase-uncertainty map computed from the total uncertainty and the predicted means \(\bar{\mu}_{N}\) and \(\bar{\mu}_{D}\).}
    \label{fig}
\end{figure}

Fig.~9 shows the uncertainty maps for a scenario with isolated objects and the corresponding wrapped-phase error. Using our model predictions, we compute wrapped phases at three modulation frequencies and compare them with the ground truth to obtain the wrapped-phase error maps shown in Fig.~9(b). We use these error maps to verify that the uncertainty maps identify regions where predictions may exhibit increased error. In Fig.~9(c) and Fig.~9(d), regions with relatively large wrapped-phase error coincide with high total uncertainty. The same trend is observed for the phase uncertainty in Fig.~9(e). In summary, the total uncertainty aggregates error sources and remains statistically calibrated. Relying solely on data uncertainty can lead to overconfidence in out-of-distribution or underfitted regions, whereas relying solely on model uncertainty fails to capture physical noise at low SNR. Phase uncertainty projects risk onto the final task variable, the unwrapped phase \(\phi\), and thus provides a directly usable quality measure. Data and model uncertainties remain valuable for identifying sources of error and guiding improvements. Data uncertainty reflects measurement noise and inherent ambiguity in fringe patterns, such as low contrast, saturation, or shadows. Model uncertainty arises from limited training coverage or generalization error and is often high under out-of-distribution or rarely observed surface conditions, identifying regions where predictions are less accurate due to gaps in model knowledge. Therefore, in practical applications, data and model uncertainties should be analysed alongside the total and phase uncertainties to draw more reliable conclusions.

Fig. ~9 (b) reports mean absolute errors of the wrapped phase of \(0.055~\mathrm{rad}\), \(0.025~\mathrm{rad}\), and \(0.032~\mathrm{rad}\) for the high, mid, and low frequencies, respectively, and an unwrapped phase error of \(0.055~\mathrm{rad}\). We attribute the closeness between the high frequency wrapped phase error and the final unwrapped phase error to the small errors at the low and mid frequencies, which do not propagate during unwrapping because these frequencies are used primarily to determine the fringe order. In addition, the wrapped phase at the medium frequency shows a smaller phase error than those at the high and low frequencies. This behaviour can also be observed across all the uncertainty maps. This is mainly because high-frequency fringes are attenuated by the system MTF, reducing modulation contrast, whereas low-frequency fringes are more susceptible to ambient illumination, non-uniform lighting, and spectral leakage \cite{takeda1983fourier}. Consistently, the estimated uncertainty is positively correlated with the overall prediction error.  

Further, if considered in a quality control setting, this experiment would provide a typical method of how the HSURE-Net allows for making better decisions. For example, in additive manufacturing, deep learning-based FPP are used for surface defect detection because of their speed and accuracy. One may face the risk of incorrectly classifying a region of the part as a defect due to the prediction failure or missing a defect. The solution is to check the estimated uncertainty maps that can reflect the possible problems of the predictions and model. This increases the reliability of the deep learning model and provides indicative metrics for people to identify the defect instead of blindly believing
\begin{figure}[htbp]
    \centering
    \includegraphics[width=\textwidth]{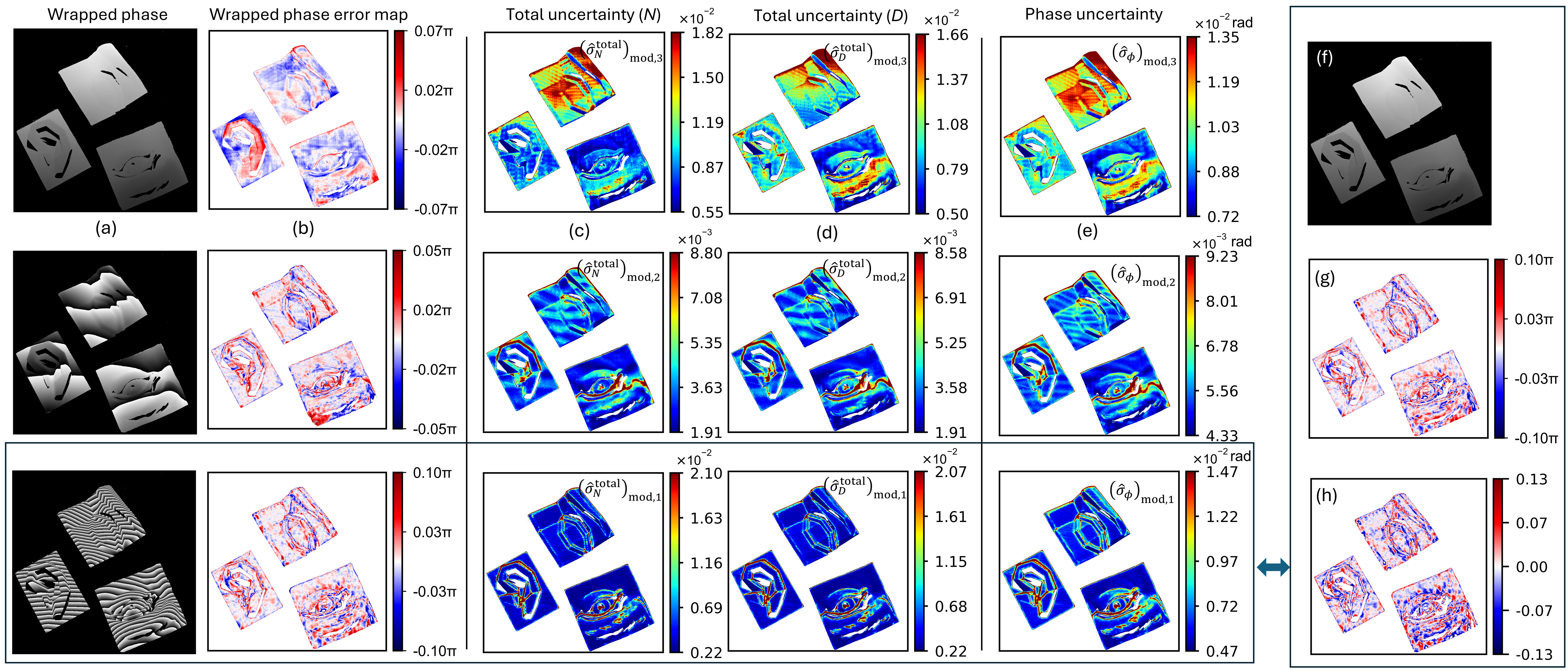}
    \caption{Association between the total and phase uncertainties and the wrapped-phase error. Uncertainty maps indicate regions where the wrapped phase derived from the predicted \(N\) and \(D\) may exhibit increased error. (a) Wrapped-phase values at three modulation frequencies \(f_{\mathrm{mod}_1}\), \(f_{\mathrm{mod}_2}\), and \(f_{\mathrm{mod}_3}\). (b) Wrapped-phase error relative to the ground truth for the maps in (a). (c) and (d) Total-uncertainty maps for \(N\) and \(D\). (e) Phase-uncertainty map.}
    \label{fig}
\end{figure}
To evaluate the dynamic measurement performance of HSURE-Net, the projector continuously projected the pattern while the camera captured images of objects in motion. Two objects were measured simultaneously within the same FOV as an in-scene control: one object remained static, whereas the other underwent rotation and translation relative to the static object. With the camera operating at 17\,fps, the frame period was set to 67\,ms. A sequence of 27 frames was acquired, corresponding to a total acquisition time of 1809\,ms. Six frames captured at time instants $\tau_k$ $(k=1,2,\ldots,6)$ were selected (Fig.~10(a)), and the corresponding reconstructions and phase uncertainty maps are shown in Fig.~10(b) and Fig.~10(c), respectively. As shown in Fig.~10(b), the static object ("lip gypsum") located at the bottom of the FOV exhibits nearly identical reconstructions across time, with a mean height MAE of 0.035\,mm. For the rotating object ("eye gypsum''), HSURE-Net still produces accurate, high-resolution reconstructions under continuous motion, achieving a mean height MAE of 0.039\,mm. The reconstruction at $\tau_4$ is partially missing due to finger occlusion during rotation, which introduces shadows in the captured image. The phase uncertainty maps in Fig.~10(c) indicate regions that are more likely to contain reconstruction errors. As the object rotates, the uncertainty of the same surface region varies slightly when it appears at different spatial locations in the image, mainly because the fringe quality changes with position. Nevertheless, the uncertainty remains primarily correlated with surface geometric complexity. Notably, when the object motion is too fast relative to the camera frame rate, motion blur artifacts may occur, which degrades reconstruction accuracy.
\begin{figure}[htbp]
    \centering
    \includegraphics[width=\textwidth]{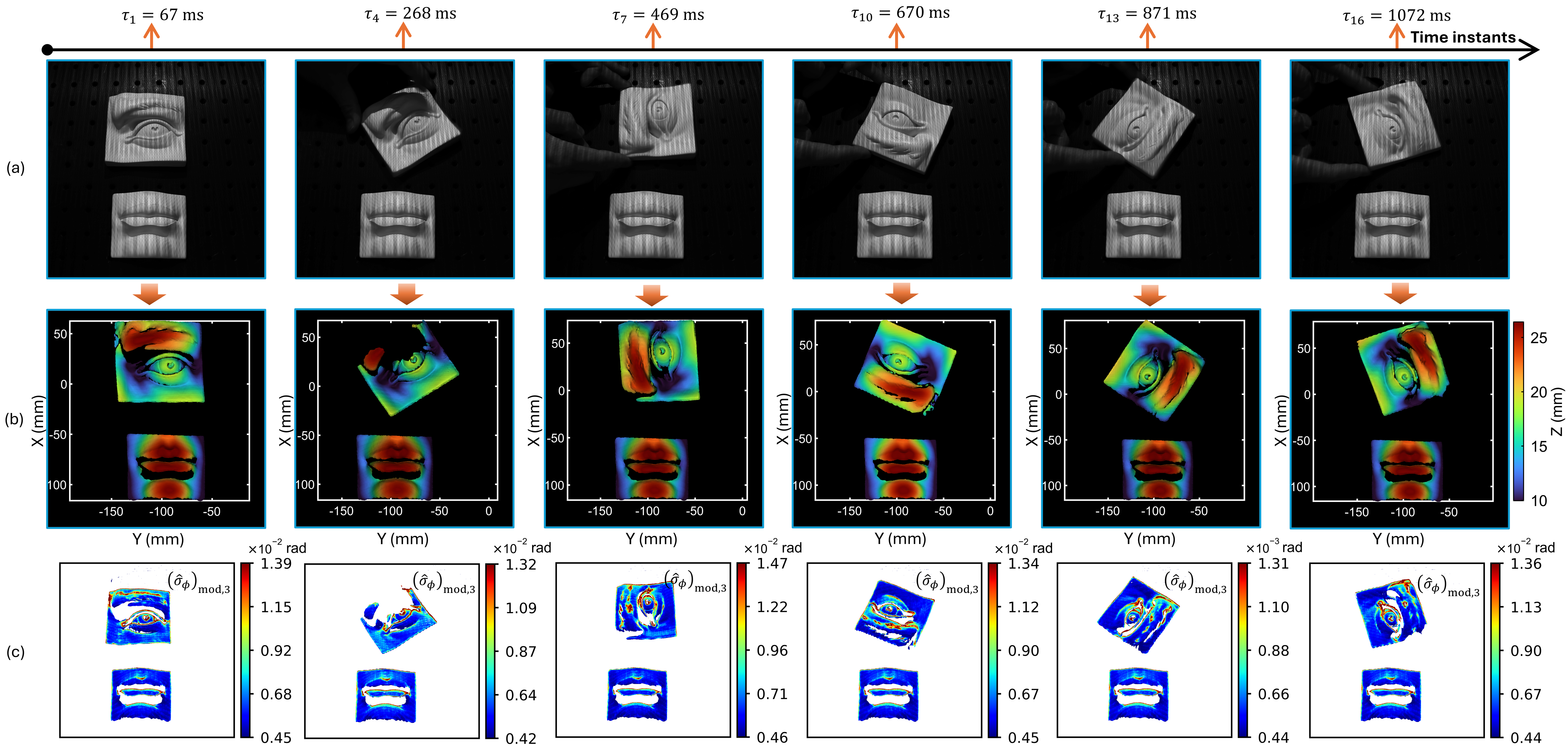}
    \caption{Static-dynamic validation of HSURE-Net within a single FOV: one object remains stationary while the other rotates relative to it. Along with the 3D reconstructions, the corresponding phase uncertainty maps are provided. (a) HFCF frames captured at time instants $\tau_k$ $(k=1,2,\ldots,6)$. (b) 3D reconstructions at each time instant. (c) Phase-uncertainty maps associated with the height maps of the corresponding reconstructions.}
    \label{fig}
\end{figure}

\section{Conclusion}

We propose a real-time 3D measurement method, termed HSU-CFPP, based on a heteroscedastic snapshot-ensemble deep neural network. It achieves high-accuracy single-frame 3D reconstruction and simultaneously outputs pixel-wise uncertainty estimates, including data uncertainty, model uncertainty, their total uncertainty, and the resulting phase uncertainty. Based on the principle of spatial FM, we adopt a frequency hierarchical strategy to encode the modulated fringes into the HFCF via carrier fringes, thereby providing phase information at multiple frequencies for phase unwrapping while mitigating spectral aliasing. The HFCF is fed as input to an HSU-Net network, which predicts the sine and cosine components (i.e., numerator and denominator) used to compute the wrapped phase. The absolute phase is then recovered from the wrapped phases at different frequencies derived from the HFCF predictions, enabling unambiguous single-shot phase unwrapping. HSU-Net models heteroscedasticity by predicting per-pixel log-variances for the numerator and denominator; data uncertainty is derived from these variances, and model uncertainty from the dispersion of snapshot predictions. These components reflect measurement noise and limited model knowledge, respectively. Their sum yields a total uncertainty that is empirically well-calibrated under a heteroscedastic Gaussian likelihood. Phase uncertainty, obtained via first-order error propagation, provides a quantitative measure of phase quality for confidence maps and downstream weighting. Experiments demonstrate the accuracy of the proposed method on both continuous surfaces and isolated discrete objects, and reveal a strong correlation between the uncertainty measures and both phase and height map errors, with regions of higher estimated uncertainty exhibiting larger measurement errors that propagate into the final 3D reconstruction. Effectively exploiting these uncertainty maps enables a measurable notion of confidence for both the deep-learning outputs and the reconstructed point clouds, which is critical for quality control in real-time FPP measurements, especially when ground truth is unavailable.

In addition, we believe that the proposed HSU-Net has strong potential for other pixel-wise regression tasks requiring calibrated uncertainty, such as optical interferometry, radar and remote sensing, and medical and scientific imaging. Recognising the importance of uncertainty analysis in optical metrology, our future work will move beyond network-level uncertainty to investigate the overall uncertainty budget of deep-learning-based real-time FPP systems, including the measurement environment, measurement hardware, reconstruction algorithms, and model outputs.

\bibliography{References}

%%%%%%%%%% If preparing manually:
% \begin{thebibliography}{1}
% \newcommand{\enquote}[1]{``#1''}

% \bibitem{Zhang:14}
% Y.~Zhang, S.~Qiao, L.~Sun, Q.~W. Shi, W.~Huang, L.~Li, and Z.~Yang,
%   \enquote{Photoinduced active terahertz metamaterials with nanostructured
%   vanadium dioxide film deposited by sol-gel method,}
%   {\protect\JournalTitle{Optics Express}} \textbf{22}, 11070--11078 (2014).

% \bibitem{Optica}
% {Optica}, \enquote{{Optica Publishing Group},}
%   \url{http://www.opg.optica.org}.

% \bibitem{FORSTER2007}
% P.~Forster, V.~Ramaswamy, P.~Artaxo, T.~Bernsten, R.~Betts, D.~Fahey,
%   J.~Haywood, J.~Lean, D.~Lowe, G.~Myhre, J.~Nganga, R.~Prinn, G.~Raga,
%   M.~Schulz, and R.~V. Dorland, \enquote{Changes in atmospheric consituents and
%   in radiative forcing,} in \enquote{Climate Change 2007: The Physical Science
%   Basis. Contribution of Working Group 1 to the Fourth Assesment Report of
%   Intergovernmental Panel on Climate Change,}  S.~Solomon, D.~Qin, M.~Manning,
%   Z.~Chen, M.~Marquis, K.~B. Averyt, M.~Tignor, and H.~L. Miler, eds.
%   (Cambridge University Press, 2007).

% \end{thebibliography}

\end{document}